\RequirePackage{amsmath}
\documentclass[runningheads]{llncs}

\usepackage{graphicx}
\usepackage{cite}
\usepackage{amssymb}
\usepackage{xcolor}
\usepackage{hyperref}
    \hypersetup{pdflang={en},
      pdfdisplaydoctitle,
      colorlinks,
      linkcolor=black,
      citecolor=black,
      urlcolor=black,
      filecolor=black}
\usepackage{listings}
\usepackage{verbatim}
\usepackage{cprotect}
\usepackage{float}
\usepackage{listings}

\lstdefinelanguage{ML}{
  alsoletter={*},
  morekeywords={datatype, of, if, *},
  sensitive=true,
  morecomment=[s]{/*}{*/},
  morestring=[b]"
}

\lstdefinelanguage{tptp}{
  alsolanguage=Prolog,
  morekeywords={axiom, conjecture},
  keywordstyle={\color{blue}},
  columns=flexible,
  basicstyle={\small\ttfamily}
}

\lstdefinelanguage{console}{
  keywords={}
}

\lstdefinelanguage{scala}{
  alsoletter={@,=,>},
  morekeywords={end,abstract, case, catch, class, def, do, else, extends, false, final, finally, for, if, implicit, import, lazy, match, new, null, object,
override, package, private, protected, requires, return, sealed, super, then, this, throw, trait, try, true, type, val, var, while, with, yield, assert, require, decreases, ensuring,=>},
  sensitive=true,
  morecomment=[l]{//},
  morecomment=[s]{/*}{*/},
  morestring=[b]"
}

\lstdefinelanguage{isabelle}{
  basicstyle=\tiny,
  alsoletter={@,=,>},
  morekeywords={lemma,proof,show,have,also,finally,qed,next,assume,finally,fix,by}
  sensitive=true,
  morecomment=[s]{(*}{*)},
  morestring=[b]"  
}

\newcommand{\codestyle}{\sffamily}

\newcommand{\smartparagraph}[1]{\textbf{#1}.\ }

\usepackage{subfiles} 

\begin{document}
\title{Verifying a Realistic Mutable Hash Table}
\subtitle{Case Study}
%
%
\author{Samuel Chassot\orcidID{0009-0000-9751-9252} \and
Viktor Kun\v{c}ak\orcidID{0000-0001-7044-9522}} 
%
\authorrunning{S.~Chassot \and V.~Kun\v{c}ak}
%
\institute{EPFL, Switzerland \\
\email{samuel.chassot@epfl.ch},
\email{viktor.kuncak@epfl.ch}}
\maketitle              
\begin{abstract}
In this work, we verify the mutable \verb|LongMap| from the Scala standard library,  a hash table using open addressing within a single array, using the Stainless program verifier. As a reference implementation, we write an immutable map based on a list of tuples. We then show that \verb|LongMap|'s operations correspond to operations of this association list. To express the resizing of the hash table array, we introduce a new reference swapping construct in Stainless. This allows us to apply the decorator pattern without introducing aliasing. Our verification effort led us to find and fix a bug in the original implementation that manifests for large hash tables. Our performance analysis shows the verified version to be within a 1.5 factor of the original data structure. 

\end{abstract}

\graphicspath{{\subfix{./res/}}}

\section{Introduction}

With the improvements in effectiveness and expanding user base of proof assistants such as Isabelle/HOL \cite{nipkowIsabelleHOL2002} and Coq \cite{teamCoqProofAssistant2023}, we are witnessing systematic verification of many purely functional data structures. The verification of these data structures is extremely effective using those tools. In the Scala language ecosystem, such verification efforts were carried out using the Stainless verifier \cite{SystemFRFormalizeda} and its predecessor Leon \cite{madhavanContractbasedResourceVerification2017}. However, verification of mutable data structures remains more challenging. As an example for hash table validation on the JVM platform, a recent attempt \cite{deboerFormalSpecificationVerification2023} provided a proof with interactive steps and an incomplete proof based on bounded model checking for one function. We consider such efforts very valuable. At the same time, our verification led us to discover a bug that bounded model checking would have likely missed, due to the large arrays required. This illustrates the limitations of bounded checks and the need for full formal verification.

In this work, we verify a data structure from the Scala standard library: the mutable \verb|LongMap[V]|, a hash table with keys of type \verb|Long| and values of a generic type \verb|V|, implemented with open addressing (with all data stored in the arrays). We verify it using Stainless, a verification framework for a subset of Scala. This is, to our knowledge, the first verified mutable map in Scala, and the first verified hash table with open addressing and quadratic probing. Our implementation follows closely the existing mutable hash table implementation of the Scala library \cite{lampScalaLongMapStdLib}, which was implemented with efficiency in mind and withstood the test of usability. Our experience helped us further assess the use of Stainless for imperative code, following recent verification of a compression algorithm \cite{BucevKuncak22FormallyQOI} and file system components \cite{HamzaETAL22NFM}.
Our paper includes the following contributions:
\begin{enumerate}
    \item As the key result, the adaptation and full formal verification of the mutable \verb|LongMap| of the Scala standard library \cite{lampScalaLongMapStdLib} using Stainless \cite{SystemFRFormalizeda, MilovancevicKuncak23ProvingDisprovingEquivalence}; our code is available on GitHub\cite{chassotLongMapCodebase};
    \item A reference implementation of a map realized as a sorted list of tuples, along with lemmas for reasoning about such maps. We use the implementation and properties of this data structure as a specification for \verb|LongMap| and find that it supports automated and inductive reasoning better than the built-in maps of Stainless;
    \item Introduction into Stainless of an operator for swapping references, which increases the expressive power of Stainless while preserving non-aliasing, allowing us to implement the \verb|repack| method of the hash table;
    \item An evaluation of the performance of both \verb|LongMap| implementations and the mutable \verb|HashMap| of the Scala standard library, showing that the performance of the verified implementation remains competitive despite the changes introduced to simplify verification.
\end{enumerate}

\subsection{Related Work}

Hash tables have been of interest in verification from the early days of the field. Guttag \cite{DBLP:journals/cacm/Guttag77} explores the use of algebraic specifications for reasoning about hash tables, though without formal connection to executable implementations. A hash table is one of the case studies\cite{laraJahobHashtablesCodebase} in the Jahob verification system \cite{Kuncak07DecisionProceduresModularDataStructureVerification, ZeeETAL08FullFunctionalVerificationofLinkedDataStructures}. The version in Jahob does not use open addressing but separate chaining with linked list buckets. Furthermore, that case study uses, as an unverified assumption, the fact that the hash function is pure and deterministic.
The Eiffel2 project offers a collection of verified data containers, impressive by its diversity\cite{polikarpovaFullyVerifiedContainer2018a}. They implemented and verified a hash table implementation using chaining. These containers are however simpler in their implementations than what appears in Scala and Java standard libraries. We were unable to explore this collection more in depth because of the unavailability of the used tools.
De Boer et al. verified JDK's \verb|IdentityHashTable|, based on open addressing and linear probing, in their case study \cite{deboerFormalSpecificationVerification2023}. The verification was done using KeY\cite{ahrendtDeductiveSoftwareVerification2016} and JJBMC\cite{beckertModularVerificationJML2020}, both accepting JML specification. They notably did not manage to provide a deductive proof for the \verb|remove| method and one of its auxiliary methods, but instead used bounded model checking for a map of up to four elements. For the more complex methods, the KeY deductive proofs required interactive steps, up 1'655 for the \verb|put| method.
Hance et al. also proposed techniques to verify distributed systems interacting with an asynchronous environment, in particular file systems~\cite{hanceStorageSystemsAre}. In this work, they developed and verified a hash table with open addressing and linear probing in Dafny. They implemented two versions of the hash table, one using immutable data structures and one with mutable ones. This separates the functional correctness and correct heap manipulation proofs, but requires implementing the hash table twice. The authors then reimplemented it using Dafny annotations to reduce the amount of code.

\section{LongMap: from Scala Library to Stainless}

A \verb|LongMap[V]| (called \verb|LongMap| in this work) is a data structure implementing a map behavior with keys of type \verb|Long|, which denotes signed 64-bit machine integers, and values of generic type \verb|V|. The mutable \verb|LongMap| of the Scala standard library \cite{lampScalaLongMapStdLib} is a hash table with open addressing and quadratic probing.

We implement a subset of the original \verb|LongMap| interface (outlined in the Section \ref{sec:verification_specification}). The \verb|apply| function returns a default value if the key is not in the map. The \verb|remove|, \verb|update|, \verb|repack| functions return a Boolean value indicating the operation success.

The keys and values are stored in two arrays called \verb|_keys| and \verb|_values| respectively. Both are of size $N=2^n$ for some $3 \leq n \leq 30$. The index of a given key is computed using a hash function. The corresponding value is stored in the second array at the same index as its key. We define \verb|mask| $=N - 1$.

There are 2 special values in \verb|_keys|: \verb|0| and \verb|Long.MinValue|. Value \verb|0| is used to indicate a free spot while \verb|Long.MinValue| is a \textit{tombstone} value, indicating that a key was removed at this spot.

We use open addressing with quadratic probing \cite{chenQuantitativeEvaluationPersistent2023,maurerProgrammingTechniqueImproved1968} to resolve collisions. The probing function is a quadratic function of the number of collisions, recursively defined as $index_{x+1} = (index_x + 2 * (x + 1)*x - 3)\ \&\ mask$. Our verification does not depend on the details of the computation, but it checks that the implementation is pure, terminating, and free of runtime errors.

All operations rely on two elementary ones: 1) looking for a key (\verb|seekEntry|), and 2) looking for a key or an empty spot (\verb|seekEntryOrOpen|). These two operations use quadratic probing and the special values \verb|0| and \verb|Long.MinValue| in \verb|_keys|. The algorithms are outlined in the Appendix~\ref{appendix:algorithms}. For example, \verb|update(k: Long, v: V)| first computes \verb|i = seekEntryOrOpen(k)|: if \verb|k| is at index \verb|i|, it writes \verb|_values(i) = v|; if the function returns an open spot, it writes \verb|_keys(i) = k| and \verb|_values(i) = v|.

\subsection{Adapting to Stainless}
\label{sec:adaptations}
We make several changes to the original code to either comply with the supported subset of Scala or to improve the SMT solvers performance at solving queries.

\smartparagraph{Tail recursion} We replace \verb|while| loops with tail-recursive functions. Stainless can internally transform \verb|while| loops to tail-recursive functions automatically, but we have better control over specifications if we manually transform the source. The Scala compiler transforms tail-recursive functions back to loops during compilation, so no performance is lost.

\smartparagraph{Loop termination} We introduce a counter and condition that stops \lstinline|while| loops in \verb|seekEntry| and \verb|seekEntryOrOpen| when the counter is large enough. This allows us to prove that the loops terminate.

\smartparagraph{Data representation} the original implementation uses the MSBs (Most Significant Bits) of the index returned by the seeking functions to indicate whether the index points to the key, a \verb|0|, or a tombstone. We replace this with case classes for better code readability and improved verification experience, as bitwise operations are often slow in SMT solvers.

\smartparagraph{Typing and initialization of arrays} The array \verb|_values| contains values of type \verb|trait ValueCell[T]| with two implementations: \verb|ValueCellFull[T](t: T)| and \verb|EmptyCell[T]| because Stainless does not support \verb|null|s, and the \verb|Array.fill| function does not support generically typed arrays. In the original implementation, \verb|_values| is an \verb|Array[AnyRef]|, containing \verb|null| by default, and using casts to store and access values.

\smartparagraph{Refactoring} We extract common behavior of \verb|seekEntry| and \verb|seekEntryOrOpen| into another function, to reduce code and verification redundancy. Next, we move most of the functions to a companion object. Proving the preservation of the invariants about all the class' attributes for each method, even those not modified or read, is very costly. By moving functions to a companion object, only the manipulated attributes are present, simplifying proof effort. 
Finally, we split the implementation into two classes, following the decorator design pattern, as detailed in Section~\ref{sec:decorator}.

\section{Specification and Verification}
\label{sec:verification_specification}

To write the specification of the mutable \verb|LongMap|, we first implement \verb|ListLongMap|, an immutable map backed by a strictly ordered list of pairs \verb|(Long, V)|. 
The signatures and specifications of \verb|ListLongMap| methods are in the Appendix~\ref{appendix:listLongMap}.
We specify the mutable \verb|LongMap|, ensuring it behaves as the corresponding \verb|ListLongMap|. We provide a ghost method (not executed at runtime) in \verb|LongMap|, which returns an instance of \verb|ListLongMap|. Figure \ref{fig:mutableLongMapInterface} shows the \verb|LongMap| interface and specification. Table \ref{tab:loc} shows the lines of code for program, specification and proofs for both maps. The method \verb|valid| is the data structure representation invariant (Appendix~\ref{appendix:repInvariant}) stating, among others, that the inserted elements can be found when searching subsequently using the same probing function.
\begin{figure}
    \centering
\begin{lstlisting}[language=scala]
def contains(key: Long): Boolean = { ...
} ensuring (res => valid && (res == map.contains(key)))
def apply(key: Long): V = { ...
} ensuring (res => valid && 
                   (if (contains(key)) res == map.get(key).get
                    else res == underlying.v.defaultEntry(key)))
def update(key: Long, v: V): Boolean = { ...
} ensuring (res => valid && 
    (if (res) map.contains(key) && (map == old(this).map + (key, v)) 
     else map == old(this).map))
def remove(key: Long): Boolean = { ...
} ensuring (res => valid && (if (res) map == old(this).map - key 
                             else map == old(this).map))
def repack(): Boolean = { ... } ensuring (res => !res || map == old(this).map)
\end{lstlisting}
\vspace*{-2mm}
    \caption{Mutable LongMap interface and specification (note that we omit preconditions in this figure    : they are only \textit{valid} invariant checks, if any).
    \label{fig:mutableLongMapInterface}}
\end{figure}

\subsection{Decorator Pattern for Modular Proofs}
\label{sec:decorator}

To better modularize the proof, we split the implementation into two classes, following the decorator design pattern. 
First, \verb|LongMapFixedSize| implements the \verb|LongMap| specification in \autoref{fig:mutableLongMapInterface} except \verb|repack|, with arrays of fixed length passed to the constructor.
Then, we implement \verb|LongMap| as a decorator of \verb|LongMapFixedSize|. It implements the same interface and forward all operations to an underlying instance of \verb|LongMapFixedSize|. The \verb|repack| operation consists in creating a new underlying instance and inserting all the pairs. 
A key observation about the original implementation of \verb|repack| (see code in appendix \ref{appendix:repack}) is that it works in a way very similar to the \verb|update| function to insert all pairs. Only some checks are omitted because the array is assumed to be fresh so it contains no tombstone values and, initially, no keys. This observation allows us to use \verb|update| to implement the \verb|repack| method without a significant impact on the performance, while simplifying the proof.

\subsection{Swap Operation for More Expressive Unique Reference}

As discussed in Section~\ref{sec:decorator}, the \verb|LongMap| class relies on an underlying instance of \verb|LongMapFixedSize|. The underlying instance needs to be replaced by a new one during the \verb|repack| operation. 
The repacking process first computes the new mask, then creates a new instance with this size and inserts all pairs, and finally replaces the current underlying instance by this new one.
Aliasing appears when replacing \verb|this.underlying| by \verb|newInstance|, yet Stainless disallows it. We can, however, observe that there is no need for aliasing, because the reference \verb|newInstance| is not accessed after the assignment. 
We thus introduced the swap operation \cite{DBLP:journals/tse/HarmsW91} into the Stainless verifier. Stainless now has a \verb|Cell[@mutable T](var v: T)| class that encapsulates a mutable variable and offers a \verb|swap| operation to swap the content of two cells. This construct enlarges the expressiveness of Stainless without the need for aliasing, and enables implementing a resizable data structure on top of a fixed-size one.

\begin{table}[tp]
\centering
\begin{tabular}{|l|r|r|r|}
\hline
Class & Program LOC & Proof+spec. LOC & Total LOC \\
\hline
ListLongMap & 156 & 678 & 834\\
MutableLongMap & 409 & 7'358 & 7'767\\
\hline
\end{tabular}
\vspace{2mm}
\caption{Lines of code for program, as well as specification and proof. We use many ghost functions to express induction proofs. When a function has many arguments, we typically typeset each argument on a separate line, contributing to line counts.
\label{tab:loc}}
\end{table}

\subsection{Finding and Confirming a Bug in The Original Implementation}

During the verification, we discovered that the \verb|repack| operation does not satisfy the specification stated in its documentation. 
The documentation states that the map can accommodate up to $2^{30}$ values (preferably not more than $2^{29}$)\cite{lampLongMapSpecification}. However, a number of keys greater than or equal to $2^{28}$ makes \verb|repack| loop forever. The bug arises in the computation of the new \verb|mask|. The original algorithm is shown in the Appendix \ref{appendix:newMask}. The bug is an integer overflow: a mask candidate is reduced while it is \verb|> _size * 8|. However, if \verb|_size * 8| overflows, i.e., \verb|_size| $\geq 2^{28} $, the mask candidate is reduced past \verb|_size|. The new array cannot accommodate all the keys. We fix the bug by modifying the loop condition, then prove that the function always returns a large enough valid mask. The fixed implementation and its specification are in the Appendix \ref{appendix:fixedNewMask}.
Despite the small scope hypothesis~\cite{jacksonElementsStyleAnalyzing1996} and claims in~\cite{deboerFormalSpecificationVerification2023}, we do not expect that bounded model checking would have discovered this bug, given that it occurs only after inserting so many key-value pairs.

\begin{figure}[bt]
\includegraphics[width=0.5\textwidth]{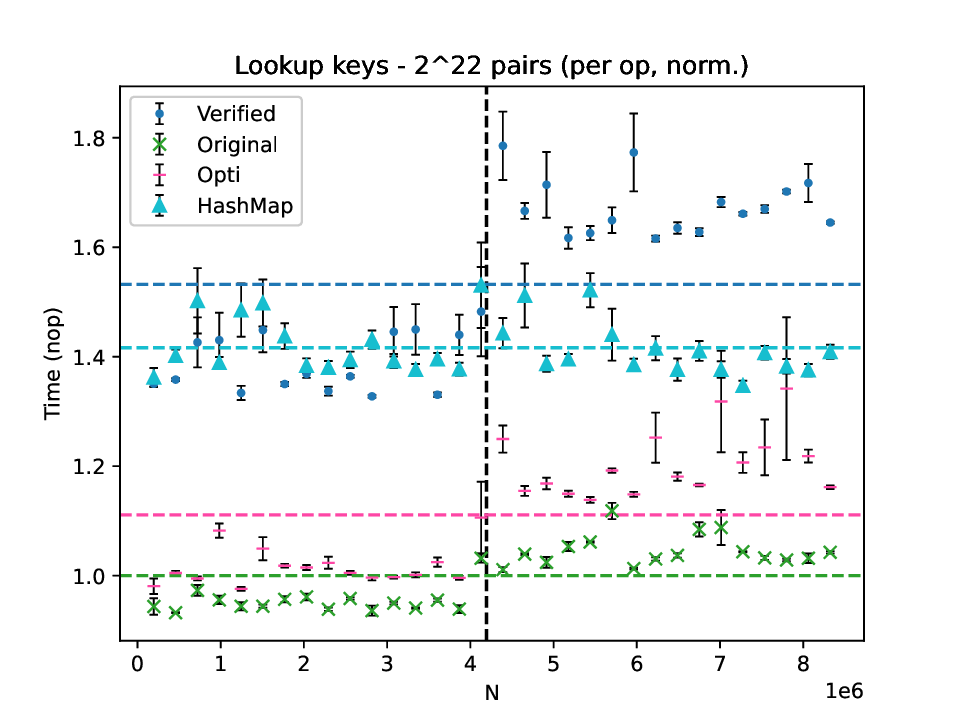}
\includegraphics[width=0.5\textwidth]{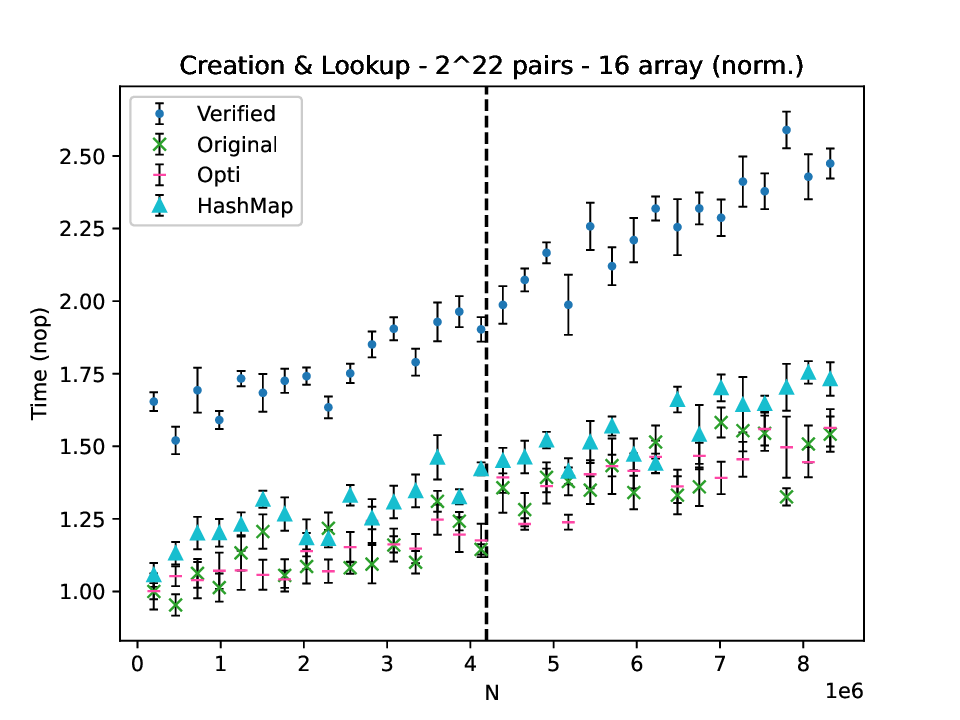}
\vspace{-5mm}
\caption{Lookup of N keys in a map prepopulated with $2^{22}$ pairs (left) (time normalized per operation) and insertion of $2^{22}$ pairs starting with an initial capacity of 16 followed by lookup of N keys (right). Horizontal lines show the average. The black vertical lines show $2^{22}$. The error bars show the 95\% CI. The time on y-axis is normalized with respect to the original map.} 
\label{fig:lookups_and_creation_lookups}
\vspace{-3mm}
\end{figure}

\section{Evaluation}

We run the benchmarks on an Ubuntu 20.04.6 LTS server, with an Intel(R) Xeon(R) CPU E5-2680 v2 @ 2.80GHz, 64GB RAM. Verification takes 427 seconds when running from scratch (103 seconds when re-running with a populated verification-condition cache~\cite{GuilloudETAL23FormulaNormalizationsVerification}).

We compare the performance of our verified implementation to the original implementation~\cite{lampScalaLongMapStdLib}, the general \verb|HashMap| of the Scala standard library\cite{lampHashMapScalaStandard}, and an optimized version of the verified implementation (denoted \verb|Opti|). We use \verb|Long| as the type of stored values.
We consider three scenarios: looking up keys in a prepopulated map, populating the map first then looking up keys, and populating the map where some keys are removed and inserted back, before looking up keys (see Appendix~\ref{appendix:benchmarkScenarios}). Results are shown in \autoref{fig:lookups_and_creation_lookups}; further results are in the Appendix~\ref{appendix:mapBenchmark}.
Our verified \verb|LongMap| is $1.5\times$ slower than the original implementation in lookups only, see~\autoref{fig:lookups_and_creation_lookups} (left). The performance gap is similar when taking the population process into account, see~\autoref{fig:lookups_and_creation_lookups} (right). We argue that this falls within an acceptable margin. The gap is bigger when more repack operations happen, as shown by the results of the benchmarks with \verb|remove| invocations (Appendix~\ref{appendix:arrayBenchmark}).

One of the potentially undesirable aspects of the verified implementation is the pointer indirection in the \verb|_values| array (Section \ref{sec:adaptations}). To better understand its impact, we modified our verified implementation to use \verb|Array[AnyRef]| like the original (without preserving the proof). The results are shown as \verb|Opti| in all figures. This shows that this indirection is responsible for most of the overhead.
The second undesirable aspect is the fact that the creation of the \verb|_values| and \verb|_keys| arrays relies on \verb|Array.fill|. This function writes all values, which is slower than constructing an array of \verb|null|s like in the original implementation. The verified repack operation is therefore slower than the original, as it is shown by the results in the Appendix \ref{appendix:mapBenchmark}, especially Figure~\ref{fig:create_remove_lookup_start_16}. To better evaluate the impact on performance, we implement the \verb|efficientFill| function that takes a called-by-value parameter instead of called-by-name like \verb|Array.fill|, and compare these two to the \verb|Array| constructor and the \verb|Arrays.fill| Java function. Results are in the Appendix \ref{appendix:arrayBenchmark}, showing that the difference is significant for the \verb|_values| array. Nonetheless, calls to \verb|repack| are infrequent, so this performance loss should be limited in practice.
The final aspect is the way in which the seek functions pass information to the caller in MSBs of the index or as case classes (Section~\ref{sec:adaptations}). This has a limited impact on the overall performance, as witnessed by the \verb|Opti| implementation.


\section{Conclusion}

We verified \verb|LongMap| from the Scala standard library, a mutable hash table with \verb|Long| keys, using open addressing and quadratic probing. This work led us to identify and fix a bug in the original library implementation. To support verification, we introduced a mutable cell with a \verb|swap| operation, improving expressiveness without introducing aliasing. The performance evaluation of our verified implementation against the original shows a slowdown of around $1.5\times$, which we believe to be acceptable. The changes we needed to perform point to directions for further improving verification support for efficient Scala constructs.

%
%
%
%
\clearpage
\bibliographystyle{splncs04}
\bibliography{references}

\newpage
\appendix
\section{Performance Evaluation Scenarios}
\label{appendix:benchmarkScenarios}
The three performance evaluation scenarios we consider are the following:
\begin{enumerate}
    \item Lookup N keys in a random order
    \item Populate the map and then lookup N keys in a random order, starting with 2 different initial size of array
    \item Populating process with remove and then lookup N keys in a random order, starting with an initial array of $2^{17}$ elements
    
    Populating the map consists in inserting all keys ($2^{15}$ or $2^{22}$ depending on the version) while the populating process with remove consists in 
    \begin{enumerate}
        \item Inserting all key-value pairs ($2^{15}$ or $2^{22}$)
        \item Removing half of them
        \item Inserting all of them again
    \end{enumerate}
\end{enumerate}

and for each, we run it for two number of key-value pairs:
 \begin{enumerate}
    \item $2^{15}$, with data points with $N \in \{1024*i \text{ for } 0\leq i \leq 64\}$ lookups.
    \item $2^{22}$, with data points with $N \in \{131072*i \text{ for } 0\leq i \leq 64\}$ ($131072 = 2^{17}$) lookups.
\end{enumerate}

\section{Map Benchmark Results}
\label{appendix:mapBenchmark}

\begin{figure}[H]
\includegraphics[width=0.5\textwidth]{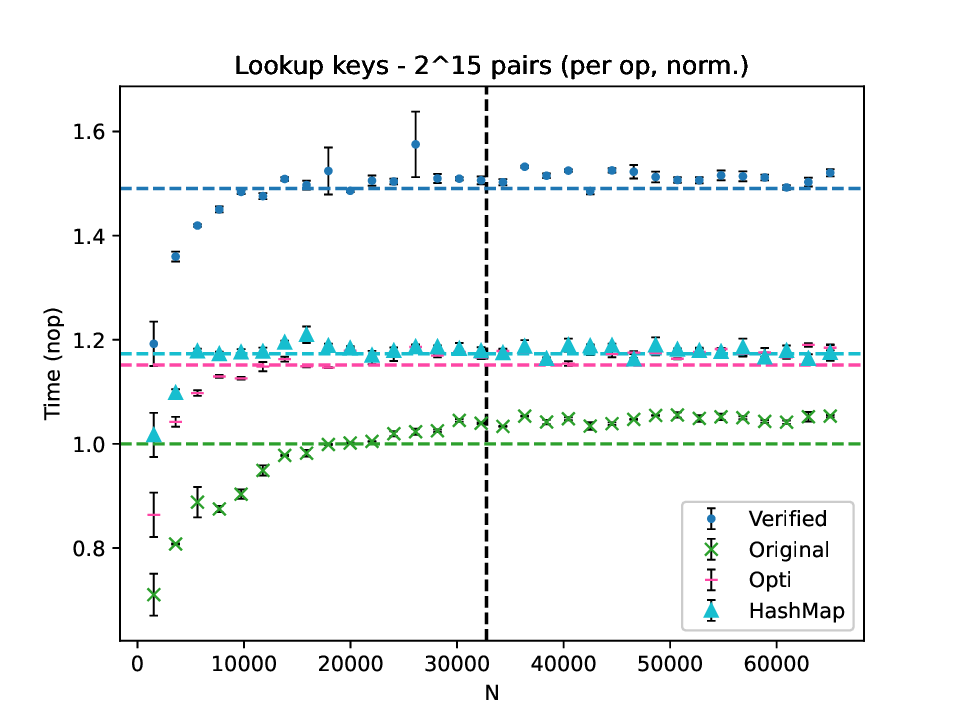}
\includegraphics[width=0.5\textwidth]{Lookup_keys_2to22_pairs_per_op_norm..eps}
\caption{Lookup of N keys in a prepopulated map with $2^{15}$ pairs (left) (time normalized per operation) and $2^{22}$ pairs (right) (time normalized per operation). Horizontal lines show the average. The black vertical line shows $2^{15}$ (left) and $2^{22}$ (right). The error bars show the 95\% CI. The time on y-axis is normalized with respect to the original map.} 
\label{fig:lookup_norm}
\end{figure}

\begin{figure}[H]
\includegraphics[width=0.5\textwidth]{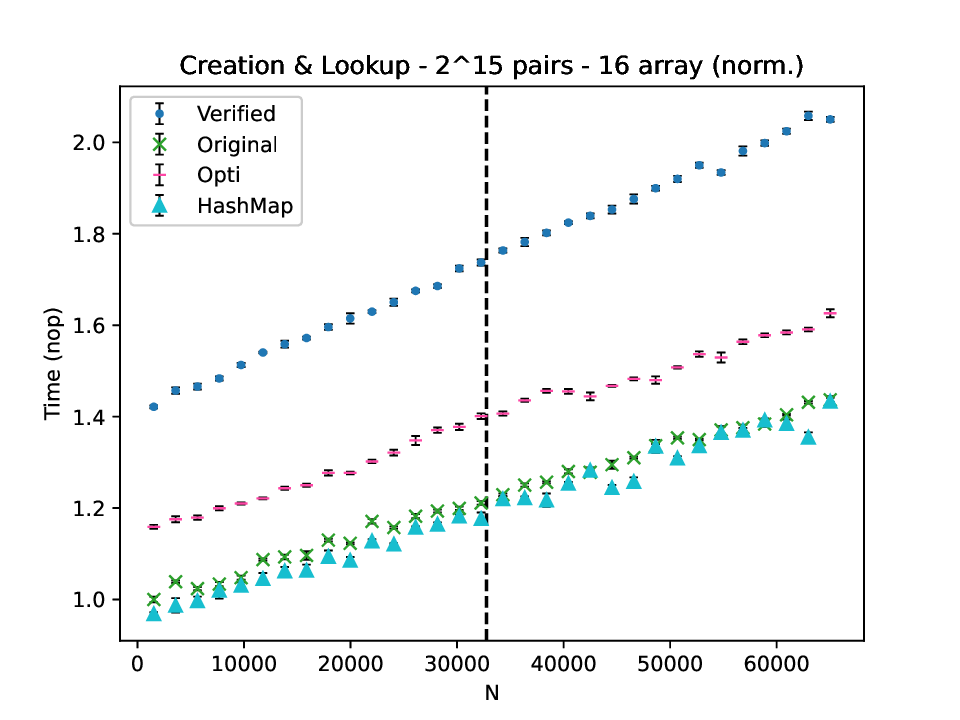}
\includegraphics[width=0.5\textwidth]{Creation_and_Lookup_2to22_pairs_16_array_norm..eps}
\caption{Insertion of $2^{15}$ pairs (left) and $2^{22}$ pairs (right) starting with an initial capacity of 16 followed by lookup of N keys. The black vertical line shows $2^{15}$ (left) and $2^{22}$ (right). The error bars show the 95\% CI. The time on y-axis is normalized with respect to the original map.}
\label{fig:create_lookup_start_16}
\end{figure}

\begin{figure}[H]
\includegraphics[width=0.5\textwidth]{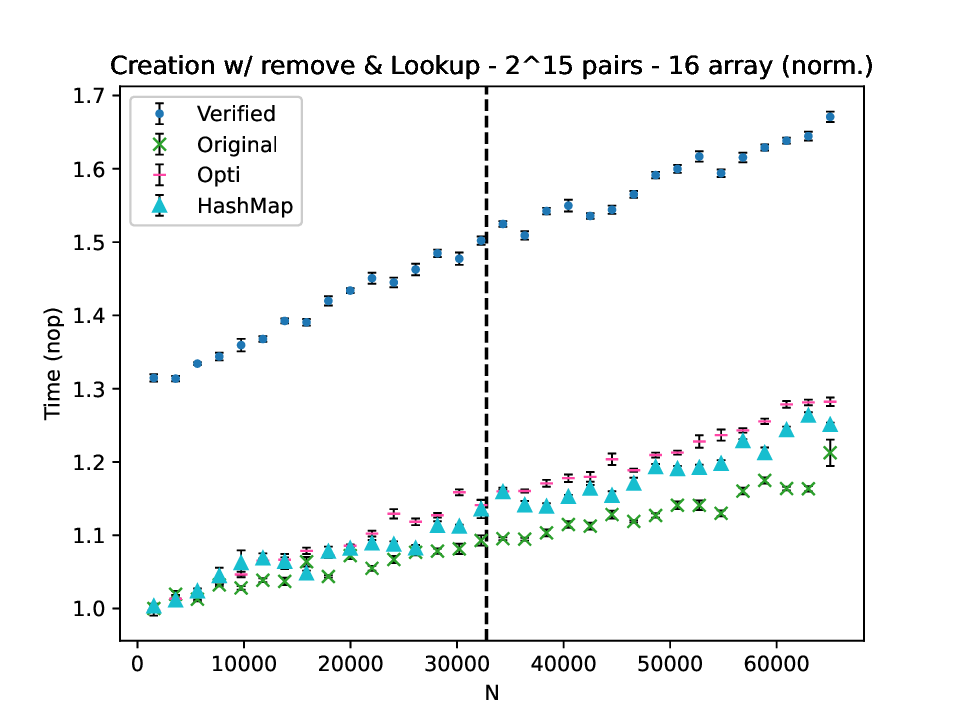}
\includegraphics[width=0.5\textwidth]{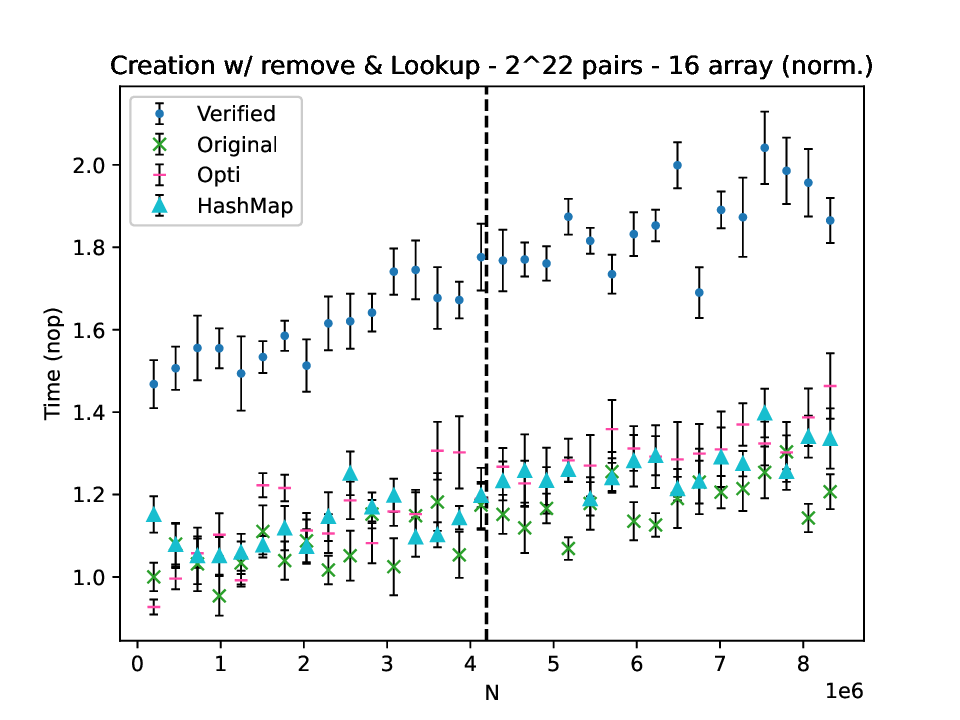}
\caption{Insertion of $2^{15}$ pairs (left) and $2^{22}$ pairs (right) starting with an initial capacity of 16 followed by lookup of N keys. During the insertion process, all pairs are inserted, then one half is removed, then all pairs are inserted again. The black vertical line shows $2^{15}$ (left) and $2^{22}$ (right). The error bars show the 95\% CI. The time on y-axis is normalized with respect to the original map.} 
\label{fig:create_remove_lookup_start_16}
\end{figure}

\begin{figure}[H]
\includegraphics[width=0.5\textwidth]{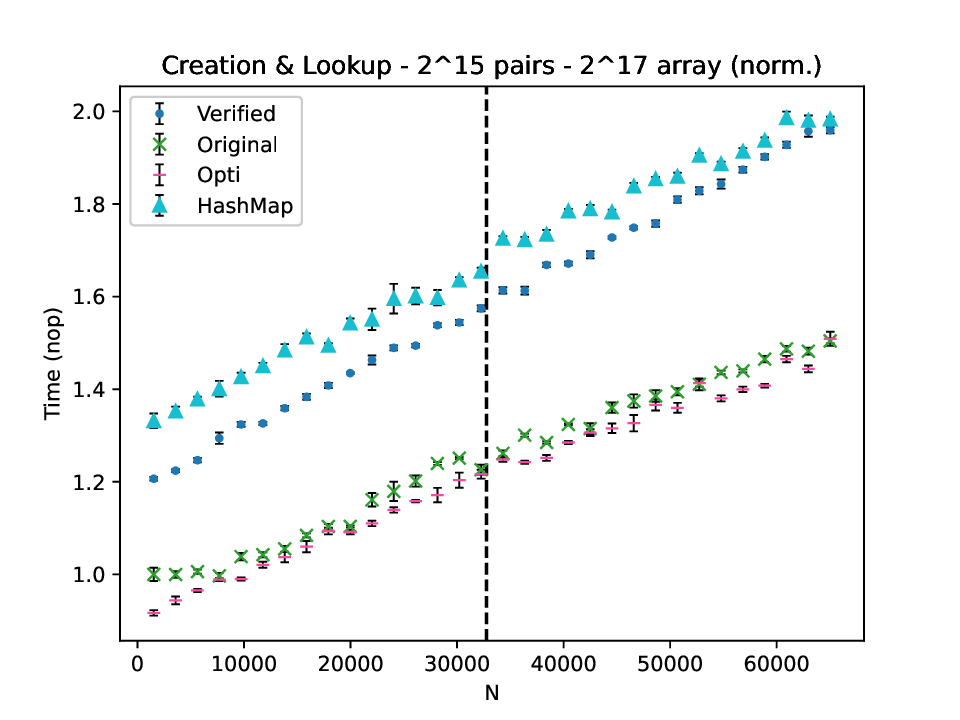}
\includegraphics[width=0.5\textwidth]{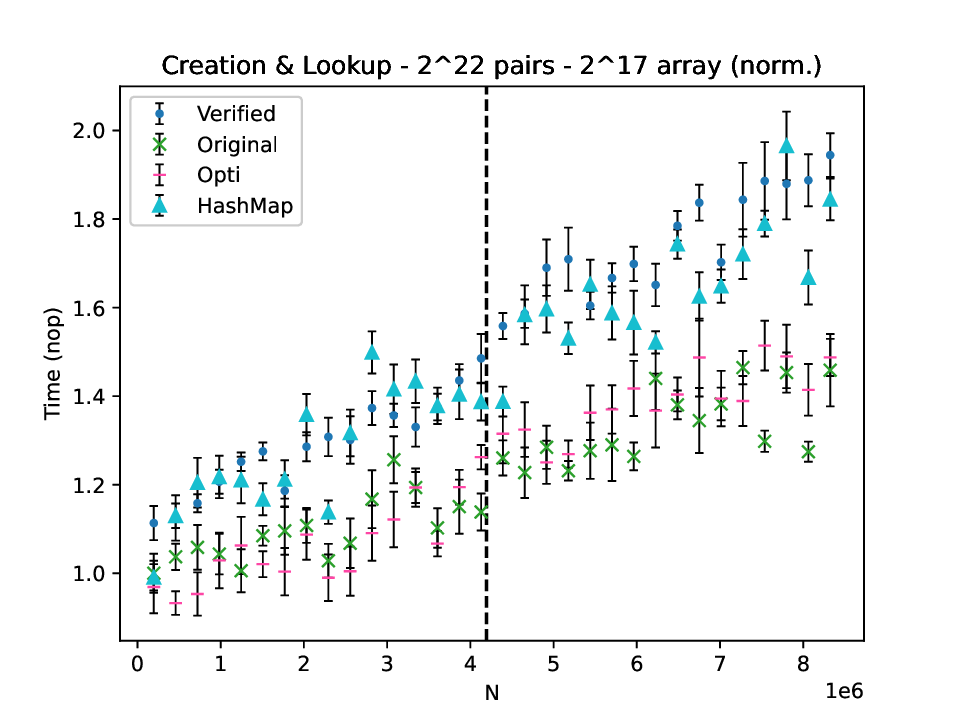}
\caption{Insertion of $2^{15}$ pairs (left) and $2^{22}$ pairs (right) starting with an initial capacity of $2^{17}$ followed by lookup of N keys. The black vertical line shows $2^{15}$ (left) and $2^{22}$ (right). The error bars show the 95\% CI. The time on y-axis is normalized with respect to the original map.}
\label{fig:create_remove_lookup_start_2to17}
\end{figure}

The figures \ref{fig:lookup_norm}, \ref{fig:create_lookup_start_16}, \ref{fig:create_remove_lookup_start_16}, and \ref{fig:create_remove_lookup_start_2to17} show benchmark results for the different scenarios and map sizes.

\section{Array Benchmark Results}
\label{appendix:arrayBenchmark}

\begin{figure}[H]
\includegraphics[width=0.5\textwidth]{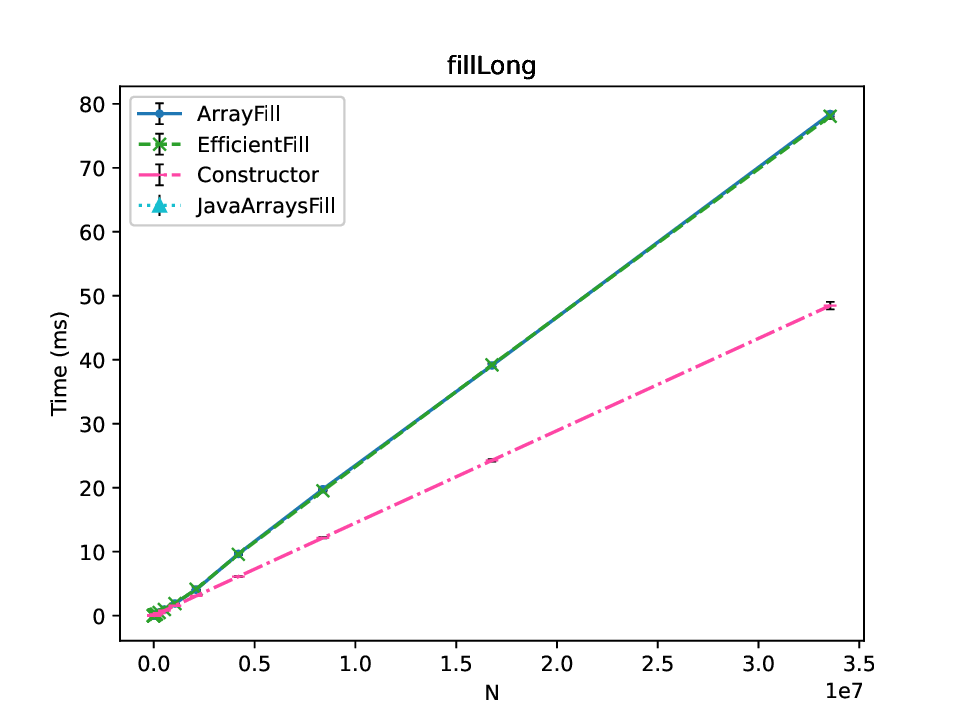}
\includegraphics[width=0.5\textwidth]{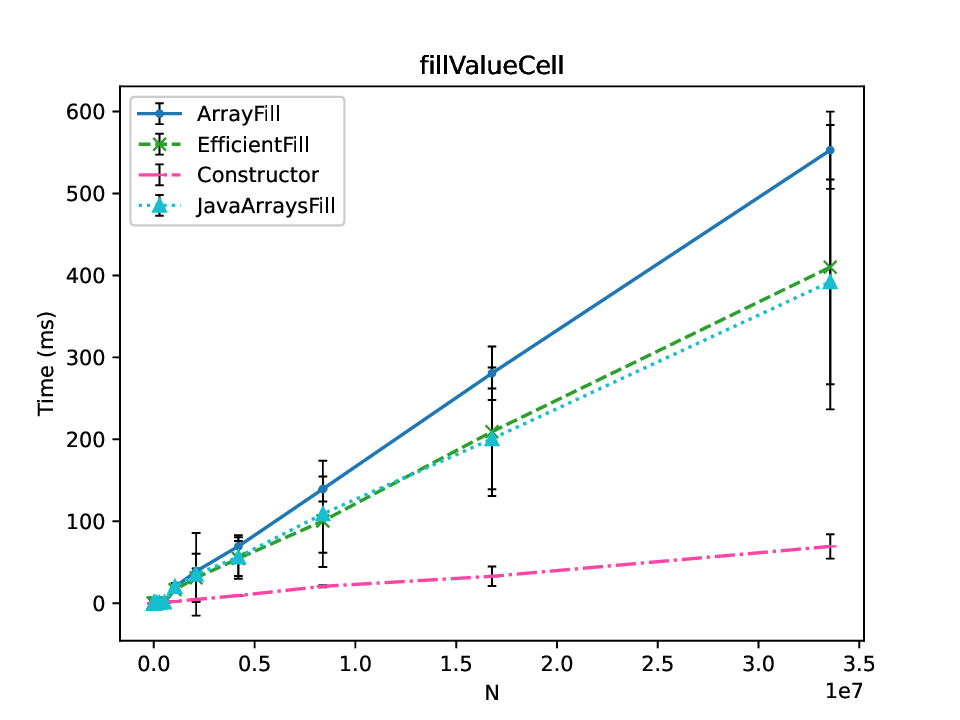}
\caption{Time to create an array of Long (left) or ValueCell (right) with Array.fill, our custom fill method, the Java Arrays.fill method, and the Array constructor.}
\label{fig_arrays_creation}
\end{figure}

Figure \ref{fig_arrays_creation} shows the performance of the different implementations of the \verb|fill| function on arrays.

\section{Key Seeking Operations Algorithms}
\label{appendix:algorithms}

This appendix presents the detailed algorithms for the \verb|seekEntry| and \verb|seekEntryOrOpen| procedures.

\verb|def seekEntry(k: Long)| algorithm (looks for a key \verb|k| in the array):

\begin{enumerate}
    \item Compute the index $i = f(k) \mod N$.
    \item If \verb|_keys(i) == k|: return \verb|i|.
    \item If \verb|_keys(i) == 0|: the key is not in the map.
    \item Otherwise, compute $i=index_{x+1}$ and go to (2).
\end{enumerate}

\verb|def seekEntryOrOpen(k: Long)| algorithm (looks for a key \verb|k| or a free spot in \verb|_keys|):

\begin{enumerate}
    \item Compute the index $i = f(k) \mod N$.
    \item If \verb|_keys(i) == k|: return \verb|i|.
    \item If or \verb|_keys(i) == 0|: return \verb|iVacant| if defined, or \verb|i|.
    \item If \verb|_keys(i) == Long.MinValue|: the spot is free, but the key might appear later in the array. Indeed, if \verb|k| collisions with the key that stood here, it would appear later. If \verb|k| is not found later, \verb|i| is the preferred free spot to return, so it is stored as \verb|iVacant = i|.
    \item Otherwise, compute $i=index_{x+1}$ and go to (2).
\end{enumerate}

To retrieve the value for a key $k$, it first computes \verb|i = seekEntry(k)|: if \verb|k| is not in the map, it returns the default value, otherwise, it returns \verb|_values(i)|.

To remove a key $k$, it computes \verb|i = seekEntry(k)|: if \verb|k| is not in the map, it does nothing, otherwise, it writes \verb|_keys(i) = Long.MinValue| and \verb|_values(i) = 0|.

\section{Original Repack Operation Implementation}
\label{appendix:repack}
\begin{lstlisting}[language=scala]
def repack(newMask: Int): Unit = {
    val ok = _keys
    val ov = _values
    mask = newMask
    _keys = new Array[Long](mask+1)
    _values = new Array[AnyRef](mask+1)
    _vacant = 0
    var i = 0
    while (i < ok.length) {
      val k = ok(i)
      if (k != -k) {
        val j = seekEmpty(k)
        _keys(j) = k
        _values(j) = ov(i)
      }
      i += 1
    }
}
def seekEmpty(k: Long): Int = {
    var e = toIndex(k)
    var x = 0
    while (_keys(e) != 0) { x += 1; e = (e + 2*(x+1)*x - 3) & mask }
    e
  }
\end{lstlisting}

\section{ListLongMap Specification}
\label{appendix:listLongMap}
This appendix presents the entire \verb|LongMap| interface and its specification.
\begin{lstlisting}[language=scala]
case class ListLongMap[B](toList: List[(Long, B)]) {
  def isEmpty: Boolean = toList.isEmpty
  def size: Int
  def contains(key: Long): Boolean = { //...
  }.ensuring(res => !res || this.get(key).isDefined)
  def get(key: Long): Option[B]
  def apply(key: Long): B = {
    require(contains(key))
    get(key).get
  }
  def +(keyValue: (Long, B)): ListLongMap[B] = { //...
  }.ensuring(res =>
    res.contains(keyValue._1) && res.get(keyValue._1) == Some[B](
      keyValue._2
    ) && res.toList.contains(keyValue)
  )
  def -(key: Long): ListLongMap[B] = { //...
  }.ensuring(res => !res.contains(key))
}
\end{lstlisting}

\section{Original New Mask Computation Function}
\label{appendix:newMask}

The procedure to compute a new mask during \verb|repack| operation in the original implementation is the following (note that some variables were renamed to match our implementation variable names):

\begin{lstlisting}[language=scala]
var m = oldMask
if (_size + _vacant >= 0.5*oldMask && 
    !(_vacant > 0.2*oldMask)) then 
        m = ((m << 1) + 1) & MAX_MASK
while (m > 8 && 8*_size < m) m = m >>> 1
// m is the new mask
\end{lstlisting}

where \verb|oldMask| is the current mask when calling \verb|repack|, \verb|_size| is the number of keys in the array (not in the map as \verb|Long.MinValue| and \verb|0| are stored differently), \verb|_vacant| is the number of tombstone values in the array and \verb|MAX_MASK| is equal to $2^{30} -1$ i.e., the biggest allowed mask.

Here is the same function slightly modified to work with Stainless. The semantics is not modified.

\begin{lstlisting}{language=scala}
def originalNewMask(mask: Int, _size: Int, _vacant: Int) = {
    require(validMask(mask))
    require(_size >= 0 && _size <= mask + 1)
    require(_vacant >= 0)
    var m = mask
    if (2 * (_size + _vacant) >= mask && 
       !(5 * _vacant > mask)) m = ((m << 1) + 1) & IndexMask
    while (m > 8 && 8 * _size < m) {
      decreases(m)
      m = m >>> 1
    }
    m
} ensuring (res => validMask(res) && _size <= res + 1) 
\end{lstlisting}

Running Stainless on it returns a counter example showing the presence of a bug:
\begin{lstlisting}{language=scala}
// [Warning ] Found counter-example:
// [Warning ]   _vacant: Int -> 10256777
// [Warning ]   mask: Int    -> 1073741823 
// [Warning ]   _size: Int   -> 603979777
\end{lstlisting}

The following program triggers the bug in the original Scala standard library implementation by looping forever. 

\begin{lstlisting}{language=scala}
import scala.collection.mutable.LongMap

/** This makes the LongMap hangs, when the size reaches 268'435'456. */
object BugLongMap {
  def triggerBug(): Unit = {
    val m = LongMap[Long]()
    for (i <- 0 until 1 << 29) {
      m.update(i, i)
      println(f"m.size = ${m.size} for i = $i")
    }
  }
}

@main def main(): Unit = {
  BugLongMap.triggerBug()
}
\end{lstlisting}

To the best of our knowledge, this error was not known previously. The library documentation claims that the hash table works up to capacity exceeding the one that triggers this infinite loop, stating \cite{lampLongMapSpecification}:

\emph{This map is not intended to contain more than $2^{29}$ entries (approximately 500 million). The maximum capacity is $2^{30}$, but performance will degrade rapidly as $2^{30}$ is approached.}

\

However, the infinite loop appears already after inserting $2^{28}$ elements.

\section{Fixed New Mask Computation Function}
\label{appendix:fixedNewMask}

The following function is the procedure that computes the new mask during the \verb|repack| process, revised to ensure that the mask is valid.

\begin{lstlisting}[language=scala]
def computeNewMask(oldMask: Int, _vacant: Int, _size: Int): Int = {
  require(validMask(oldMask))
  require(_size >= 0 && _size <= oldMask + 1)
  require(_vacant >= 0)
  var m = oldMask
  if (2 * (_size + _vacant) >= oldMask && !(5 * _vacant > oldMask)) {
    m = ((m << 1) + 1) & MAX_MASK
  }
  while (m > 8 && 8 * _size < m && 
        ((m >> 1) & MAX_MASK) + 1 >= _size) {
    decreases(m)
    m = m >>> 1
  }
  m
} ensuring (res => validMask(res) && _size <= res + 1)
\end{lstlisting}

\section{Hash Table Representation Invariant}
\label{appendix:repInvariant}

The following is the representation invariant we use as part of the specification of the hash table. In addition to operation specific conditions, the invariant is required and ensured after all external operations of the hash table. It is defined using several auxiliary functions, some of them recursive, such as \lstinline|arrayForallSeekEntryOrOpenFound| that we also include.

\begin{lstlisting}[language=scala]
def valid: Boolean =
  simpleValid &&
  arrayCountValidKeys(_keys, 0, _keys.length) == _size &&
  arrayForallSeekEntryOrOpenFound(0)(_keys, mask) &&
  arrayNoDuplicates(_keys, 0)

def simpleValid: Boolean =
  validMask(mask) &&
  _values.length == mask + 1 &&
  _keys.length == _values.length &&
  _size >= 0 &&
  _size <= mask + 1 &&
  size >= _size &&
  size == _size + (extraKeys + 1) / 2 &&
  extraKeys >= 0 &&
  extraKeys <= 3 &&
  _vacant >= 0


def arrayForallSeekEntryOrOpenFound(i: Int)
                          (implicit _keys: Array[Long], mask: Int): Boolean =
  require(validMask(mask))
  require(_keys.length == mask + 1)
  require(i >= 0)
  require(i <= _keys.length)
  decreases(_keys.length - i)

  if (i >= _keys.length) true
  else if (validKeyInArray(_keys(i)))
    lemmaArrayContainsFromImpliesContainsFromZero(_keys, _keys(i), i)
    LongMapFixedSize.seekEntryOrOpen(_keys(i))(_keys, mask) == Found(i) &&
    arrayForallSeekEntryOrOpenFound(i + 1)
  else arrayForallSeekEntryOrOpenFound(i + 1)
\end{lstlisting}

\end{document}